\documentclass[journal]{IEEEtran}
\usepackage{cite}
\usepackage{amsmath,amssymb,amsfonts}
\usepackage{amsthm}
\usepackage{algorithm}
\usepackage{algpseudocode}
\usepackage{graphicx}
\usepackage{textcomp}
\usepackage{tabularx}
\usepackage{array}
\usepackage{bbm}
\usepackage[hidelinks,colorlinks=true,linkcolor=blue,citecolor=blue,urlcolor=black]{hyperref}
\usepackage{academicons}
\usepackage[nolist]{acronym}
\usepackage{comment}
\usepackage{subcaption}
\usepackage{subfiles} 
\usepackage{textgreek}
\usepackage{listings}
\usepackage{balance}
\usepackage{soul}
\usepackage{multirow}
\usepackage[T1]{fontenc}
\usepackage{cuted} 
\usepackage{relsize}
\usepackage{booktabs}
\usepackage{longtable}
\usepackage{cleveref}
\usepackage[table]{xcolor}
\usepackage{makecell}
\usepackage{pifont}
\usepackage[compatibility=false]{caption}
\newcolumntype{Y}{>{\raggedright\arraybackslash}X}

\def\BibTeX{{\rm B\kern-.05em{\sc i\kern-.025em b}\kern-.08em
    T\kern-.1667em\lower.7ex\hbox{E}\kern-.125emX}}
\AtBeginDocument{\definecolor{ojcolor}{cmyk}{0.93,0.59,0.15,0.02}}

\newsavebox{\mybox}
\begin{document}

\title{FM-OFDM: A Constant-Envelope Sensing Waveform with Phase-Differencing Receiver Processing}

\author{Amir Bouziane,~\IEEEmembership{}
        and~H\"{u}seyin~Arslan,~\IEEEmembership{Fellow,~IEEE}
\thanks{Amir Bouziane is with Electrical and Electronics Engineering, School of Engineering and Natural Sciences, Istanbul Medipol University, 34810 Istanbul, T\"{u}rkiye (e-mail: \href{bouziane.amir@std.medipol.edu.tr}{bouziane.amir@std.medipol.edu.tr}).
}
\thanks{H\"{u}seyin Arslan is with Electrical and Electronics Engineering, School of Engineering and Natural Sciences, Istanbul Medipol University, 34810 Istanbul, T\"{u}rkiye.
}
\thanks{
}
\thanks{
}
}

\markboth{change title}{A.BOUZIANE \textit{et al.}}


\maketitle
\begin{acronym}
  \acro{FR2}{frequency range 2}
  \acro{DTFT}{discrete-time Fourier transform}
  \acro{3GPP}{3rd Generation Partnership Project}
  \acro{6G}{sixth generation}
  \acro{OTFS}{orthogonal time frequency space}
  \acro{RMS}{Root Mean Square}
  \acro{AWGN}{additive white Gaussian noise}
  \acro{PSD}{power spectral density}
  \acro{BW}{bandwidth}
  \acro{AF}{ambiguity function}
  \acro{CE}{constant envelope}
  \acro{CE-OFDM}{constant-envelope orthogonal frequency-division multiplexing}
  \acro{FM-OFDM}{frequency modulated orthogonal frequency-division multiplexing}
  \acro{CFO}{carrier frequency offset}
  \acro{CP}{cyclic prefix}
  \acro{CP-OFDM}{cyclic-prefix orthogonal frequency-division multiplexing}
  \acro{PN}{phase noise}
  \acro{DAC}{digital-to-analog converter}
  \acro{DC}{direct current}
  \acro{DFT}{discrete Fourier transform}
  \acro{FFT}{fast Fourier transform}
  \acro{FMCW}{frequency-modulated continuous wave}
  \acro{RFM}{random frequency modulation}
  \acro{RSM}{range sidelobe modulation}
  \acro{MTSFM}{multitone sinusoidal frequency modulation}
  \acro{ACF}{autocorrelation function}
  \acro{LFM}{linear frequency modulated}
  \acro{ICI}{inter-carrier interference}
  \acro{IDFT}{inverse discrete Fourier transform}
  \acro{IFFT}{inverse fast Fourier transform}
  \acro{IoT}{Internet of Things}
  \acro{ISAC}{Integrated sensing and communication}
  \acro{ISI}{inter-symbol interference}
  \acro{CRB}{Cramer-Rao bound}
  \acro{MCRB}{modified Cramer-Rao bound}
  \acro{JSAC}{joint sensing and communication}
  \acro{NTN}{non-terrestrial network}
  \acro{OFDM}{orthogonal frequency-division multiplexing}
  \acro{OFDM-IM}{OFDM with index modulation}
  \acro{OFDM-DM}{OFDM with directional modulation}
  \acro{PA}{power amplifier}
  \acro{PAPR}{peak-to-average power ratio}
  \acro{QAM}{quadrature amplitude modulation}
  \acro{QPSK}{Quadrature Phase Shift Keying}
  \acro{PSK}{Phase Shift Keying}
  \acro{MF}{matched filter}
  \acro{RDM}{range--Doppler map}
  \acro{RF}{radio frequency}
  \acro{RMSE}{root mean square error}
  \acro{SNR}{signal-to-noise ratio}

\end{acronym}

\begin{abstract}
Cyclic-prefix orthogonal frequency-division multiplexing (CP-OFDM) is widely adopted as a reference waveform for integrated sensing and communication (ISAC). However, its high peak-to-average power ratio (PAPR) requires power-amplifier back-off, thereby reducing the available sensing link budget. Frequency-modulated OFDM (FM-OFDM) provides a constant-envelope signal with 0~dB PAPR and has demonstrated reliable communication performance in high-mobility scenarios, but its sensing characteristics remain largely unexplored. This paper characterizes FM-OFDM as a sensing waveform based on its bandwidth, ambiguity function, sidelobe behavior, and Doppler estimation capability. The analysis shows that its data-dependent sidelobe floor is incoherent and decreases through across frame integration, whereas the corresponding CP-OFDM floor remains unchanged. Consequently, under equal occupied bandwidth and transmit power, frame-level integration reverses the single-symbol performance ordering and provides FM-OFDM with approximately 20~dB of additional dynamic range for weak-target detection. Moreover, the zero-delay cut of the FM-OFDM ambiguity function is shown to be deterministic and independent of the transmitted data realization, in contrast to linearly modulated waveforms, while a closed-form expression is derived for the sidelobe floor away from the zero-delay cut. Since the nonlinear mapping between the data symbols and time-domain samples prevents the direct application of the conventional CP-OFDM range-Doppler processing chain, a weighted phase-increment Doppler estimator is developed. The proposed estimator enables closed-form prediction of the sensing floor and its crossover point with a  minimum computational complexity.
\end{abstract}


\section{Introduction}
\label{sec:intro}
 
\IEEEPARstart{I}{}n \ac{6G} systems, sensing becomes a native network function \cite{liu2022integrated,GonzalezPrelcic_ProcIEEE_2024}. Target delay and Doppler are estimated from the same waveform that carries user data. Standardization is already active, and the IEEE~802.11bf amendment defines sensing procedures for Wi-Fi \cite{IEEE80211bf_2025}. In \ac{3GPP}, a Release~19 study item addresses \ac{ISAC} channel modeling \cite{Chen_RANRel19_2025,SP231583_ISAC_Timeline}, and a common modeling framework has been agreed in RAN1 \cite{Zhang_SCIS_ISAC_CM_2024}. The ETSI Industry Specification Group on \ac{ISAC} is defining use cases, requirements and system aspects \cite{ETSI_GR_ISAC_001}. Evaluation criteria for sensing are therefore being fixed now, and a candidate waveform must be assessed for range and velocity estimation, not only for data rate.

Waveform designs for \ac{ISAC} fall into three families \cite{9924202,wei2023integrated}. Sensing centric designs start from a radar waveform and load data onto it. This preserves the delay Doppler response and constrains the data rate. Communication centric designs start from a communication waveform, so the data rate is preserved and the delay Doppler response is constrained \cite{sturm2011waveform}. Joint designs sit between the two and try to hold both, for instance by combining an \ac{OFDM} data plane with a chirp \cite{wang2022triangular,10767774}. \ac{FM-OFDM} belongs to this third family. The subcarrier structure and the demapper are those of \ac{OFDM}, while the transmitted signal is a frequency modulated carrier, so the data plane is unchanged and only the mapping from the block to the samples is replaced.

The reference waveform in current \ac{ISAC} standardization work is \ac{CP-OFDM}, and it serves as the baseline for sensing evaluation \cite{Chen_RANRel19_2025}. Its sensing receiver is simple because the received frequency domain symbols are the product of the channel and the known data, so element wise division isolates the channel response and a two dimensional \ac{FFT} across subcarriers and symbols yields the range Doppler map \cite{sturm2011waveform}. This choice is also supported analytically, since \ac{CP-OFDM} attains the lowest average ranging sidelobe level among linearly modulated waveforms under \ac{QAM} and \ac{PSK} constellations \cite{liu2025cp}. The cost appears at the transmitter. A \ac{CP-OFDM} symbol superimposes many independent subcarriers, so its envelope approaches a complex Gaussian distribution and its \ac{PAPR} grows with the number of active subcarriers \cite{ochiai2001distribution}. The \ac{PA} must then back off by several decibels to stay linear \cite{rahmatallah2013peak,o2009new}, and the difficulty grows at millimetre wave, where output power and linearity are already constrained \cite{Kumaran_TMTT_2024}. Radarrange varies with the fourth root of the transmitted power \cite{richards2005fundamentals}, so each decibel of back off costs a quarter of a decibel of detection range. 

A waveform that carries data is random, so its correlation response changes from block to block. The analysis works with its mean. Closed forms are available for the expected squared \ac{ACF} of such signals under arbitrary modulation bases and constellations, and they separate it into a deterministic part set by the pulse shape and a data dependent floor set by the constellation \cite{liu2025uncovering}. That separation is the design parameter. Pulse shaping acts on the first part \cite{liao2025pulse} and constellation shaping on the second \cite{10685511}, and the optimality of \ac{CP-OFDM} in average ranging sidelobe level follows from the same framework \cite{liu2025cp}. All of it assumes linear modulation, where the data multiplies a fixed pulse. \ac{FM-OFDM} maps the data to the samples nonlinearly, so neither the optimality result nor the design tools carry over, and its \ac{AF} has to be derived directly.

Constant envelope multicarrier waveforms remove the back off entirely and let the amplifier run at saturation. They are established for communication links \cite{824966,thompson2008constant,hernando2022frequency}. Their cost is bandwidth expansion, so they suit high frequency bands and \ac{NTN} scenarios where bandwidth is available \cite{11421032}. The best studied member is \ac{CE-OFDM}, which maps the block directly onto the carrier phase. It has been evaluated as a radar waveform \cite{thompson2009constant,stralka2007constant}, its \ac{AF} has been characterized in closed form through generalized Bessel functions, and its parameters have been optimized against sidelobe level \cite{10149678,felton2023gradient}. Because the data sits in the absolute phase rather than in the phase increment, a residual \ac{CFO} displaces every sample and has to be estimated \cite{hernando2022frequency}. Delay Doppler waveforms such as \ac{OTFS} have been studied for the same purpose, including \ac{CE} variants \cite{hadani2017orthogonal,rou2024otfs,11130717}. \ac{FM-OFDM}, in the sense of driving the instantaneous frequency of an \ac{OFDM} block with the data, has not been evaluated for sensing, and its sensing capability is listed as an open direction \cite{11421032}. The triangular \ac{FM-OFDM} of \cite{wang2022triangular} carries the same name but combines \ac{OFDM} with linear frequency modulated chirps.

Radar has its own \ac{CE} frequency modulation family, known as \ac{RFM}, where the instantaneous frequency is driven by a stochastic process instead of a deterministic sweep \cite{blunt2020principles}. The result is a \ac{CE}, a continuous phase, and a thumbtack \ac{AF}, at the cost of sidelobes that change from pulse to pulse. A single \ac{RFM} pulse cannot reach the sidelobe level of an optimized chirp at the same time bandwidth product, because a thumbtack response spreads the same energy over the whole delay Doppler plane. Those sidelobes are incoherent across pulses, so slow time combining of $P$ waveforms suppresses them by $10\log_{10}P$ \cite{blunt2020principles,owen2024analysis}. Design effort in this family goes into shaping the frequency command itself, for instance by optimizing the Fourier coefficients of an \ac{MTSFM} waveform to place the sidelobes where they are least harmful \cite{hague2021adaptive}. \ac{FM-OFDM} differs in that the frequency command is not free. It is the \ac{OFDM} block carrying user data, so the sidelobe floor is a quantity to be derived rather than designed.

Hernando and Armada introduced \ac{FM-OFDM} as a waveform for high mobility communications \cite{hernando2022frequency}. The data drives the instantaneous frequency of the carrier, where \ac{CE-OFDM} drives the carrier phase. Both are pure phase modulations, so both have a 0~dB \ac{PAPR}. The receiver is differential, so a residual \ac{CFO} is removed without estimation.

The \ac{CP-OFDM} sensing pipeline does not transfer to \ac{FM-OFDM}. The mapping from data to samples is nonlinear, so the received spectrum is not the product of the channel and the data. The standard workarounds fail, as shown in Section~\ref{sec:classical_failure}. Matched filtering remains viable, but it leaves data dependent sidelobes and a nuisance phase in the slow time samples. The contributions of this paper are as follows.

\begin{itemize}
    \item We derive the \ac{AF} of \ac{FM-OFDM} and show that its zero delay cut is deterministic and identical for every data realization, unlike the linearly modulated case where the cut varies with the transmitted symbols. Away from the origin the floor is data dependent, and we obtain it in closed form.
    \item We show that this floor is incoherent, so it falls with the integration length while the \ac{CP-OFDM} floor does not. The lowest average ranging sidelobe result of \cite{liu2025cp} holds among linearly modulated waveforms and on a per symbol average, and neither condition covers a constant envelope waveform observed over a frame. At equal occupied bandwidth and equal transmit power this reverses the single symbol ordering and gives about $20$~dB of additional dynamic range in weak target detection.
    \item We derive the occupied bandwidth of \ac{FM-OFDM} in closed form and the bound it places on the modulation index. This fixes the operating point at which the three waveforms are compared, so the comparison rests on a measured bandwidth rather than a nominal one.
    \item We propose a weighted phase increment Doppler estimator, since the nonlinear mapping blocks the standard \ac{CP-OFDM} range-Doppler chain. Parabolic weights minimize the variance and give an \ac{RMSE} decaying as $U^{-3/2}$, against $U^{-1}$ for uniform weights, and the variance expression predicts the sensing floor and its crossover in closed form. The estimator needs no transform, no frequency grid and no phase unwrapping, and runs in $\mathcal{O}(U)$ operations.
\end{itemize}

A preliminary version of this work appeared as \cite{bouziane2025constant}, which treated sensing in the discriminator domain and used an empirical bandwidth rule. The present paper replaces both, deriving the occupied bandwidth in closed form, characterizing the \ac{AF} under matched filtering, and showing that the slow time data phase does not vanish and instead sets the sensing floor.

The rest of the paper is organized as follows. Section~\ref{sec:Syst_mod} gives the system model, derives the occupied bandwidth of \ac{FM-OFDM} in closed form, and inverts it for the modulation index. Section~\ref{sec:classical_failure} shows why the \ac{CP-OFDM} range Doppler chain does not transfer, derives the \ac{AF}, and obtains the sidelobe floor. Section~\ref{sec:receiver} develops the differential Doppler estimator, the weights that minimize its variance, and the resulting sensing floor. Section~\ref{sec:results} checks the closed forms against simulation and compares the three waveforms at equal occupied bandwidth and equal transmit power. Section~\ref{sec:conclusion} concludes.
\section{System Model}
\label{sec:Syst_mod}

Let us consider a single antenna monostatic \ac{ISAC} system. The transmitter (Tx) sends an \ac{FM-OFDM} signal that is used for both data transmission and sensing. The signal propogates to the communication user, and it is also reflected by several passive targets at different ranges. These echoes arrive at the sensing receiver (Rx). Tx and Rx are colocated and isolated well enough that the direct path from Tx to Rx can be ignored. The Rx processes the echoes to estimate the target delays and Doppler shifts.

\subsection{Transmitter Model}
\label{ssec:tx_model}
Let the unit power \ac{QAM} symbols $X[k]$ that are drawn independently with zero mean populate the frequency domain vector $\mathbf{X} = [X[0], X[1], \dots, X[N-1]]^{\mathsf{T}}$ as 
\begin{equation}
    X[k] =
    \begin{cases}
        X[k], & k_0 \le k \le k_0+N_a/2-1, \\
        X^*[N-k], & N-k_0-N_a/2+1 \le k \le N-k_0, \\
        0, & \text{otherwise},
    \end{cases}
    \label{eq:X_vec}
\end{equation}
The conjugate symmetry makes $x[n]$ real valued. With $\mathcal{K}$ denoting the active set, $|\mathcal{K}|=N_a$ active subcarriers, the signal is the unitary \ac{IDFT}
\begin{equation}
    x[n] = \frac{1}{\sqrt{N}}\sum_{k\in\mathcal{K}} X[k]\,
    e^{j2\pi kn/N}, \qquad n=0,\dots,N-1 .
    \label{eq:idft}
\end{equation}
The $1/\sqrt{N}$ scaling fixes the power of $x[n]$ and is retained in every variance expression that follows, giving
\begin{equation}
    \sigma_x^2 \triangleq \mathbb{E}\big[x^2[n]\big] = \frac{N_a}{N} .
    \label{eq:x_var}
\end{equation}
The modulator maps $x[n]$ to the instantaneous frequency
\begin{equation}
    f[n] \triangleq m\sqrt{\frac{N}{N_a}}\,x[n],
    \label{eq:freq_cmd}
\end{equation}
with $m$ the modulation index. The factor $\sqrt{N/N_a}$ cancels $N_a/N$ from \eqref{eq:x_var}.  so
\begin{equation}
    \sigma_f^2 \triangleq \mathbb{E}\big[f^2[n]\big] = m^2 .
    \label{eq:sigma_f}
\end{equation}
Integrating and exponentiating gives the transmitted signal,
\begin{equation}
    \phi[n] = 2\pi\sum_{i=0}^{n} f[i], \qquad s[n] = A_c\, e^{j\phi[n]} ,
    \label{eq:phase}
\end{equation}
with $\phi[-1]\triangleq0$, whose \ac{PAPR} is $0$~dB since $|s[n]|=A_c$ at every sample.
\begin{figure}
    \centering
    \includegraphics[width=0.8\linewidth]{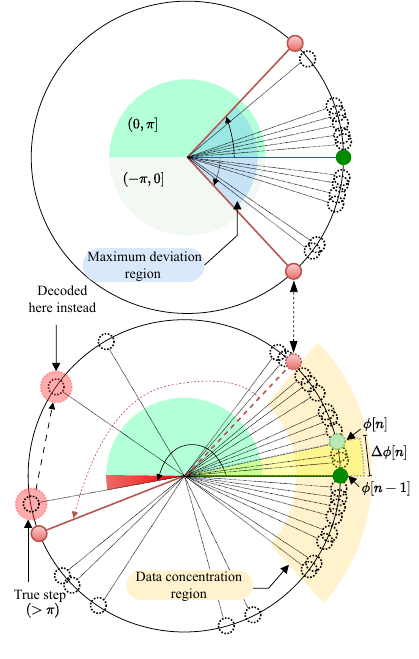}
    \caption{Phase step on the unit circle traced by $s[n]$.}
    \label{fig:mod_index}
\end{figure}

The accumulation in \eqref{eq:phase} separates \ac{FM-OFDM} from \ac{CE-OFDM}. That waveform maps $x[n]$ directly onto the phase, $\phi_{\mathrm{CE}}[n]=2\pi m\sqrt{N/N_a}\,x[n]$, with no accumulation \cite{thompson2009constant,11130717}. Here $x[n]$ instead sets the phase step, so the phase at sample $n$ depends on all preceding samples.

Differencing \eqref{eq:phase} gives $\phi[n]-\phi[n-1]=2\pi f[n]$, with standard deviation $2\pi m$. The receiver reads $f[n]$ back from this step, but only within one cycle per sample. The step is unambiguous when $|f[n]|<\tfrac12$, because $x[n]$ is a sum of $N_a/2$ independent symbols it is well approximated as Gaussian \cite{ochiai2001distribution}, and the model gives the phase step no upper bound, so the deterministic peak condition of \cite[Eq.~7]{hernando2022frequency} is relaxed here to a probabilistic one. A wrap occurs when the step exceeds $\pi$, so
\begin{equation}
    P_w = 2\,Q\!\left(\frac{1}{2m}\right) ,
    \label{eq:wrap_prob}
\end{equation}
with $Q(\cdot)$ the Gaussian tail probability, falling rapidly as $m$ decreases.

Fig.~\ref{fig:mod_index} showcases the two \ac{CE} mappings on the unit circle traced by $s[n]$. \ac{CE-OFDM} places the data at the dot positions $\phi[n]$ themselves, where a residual \ac{CFO} displaces every dot by a growing amount and must be estimated and corrected. \ac{FM-OFDM} places the data in the gap $\Delta\phi[n]$ between consecutive dots, so a \ac{CFO} adds the same increment to every gap and leaves the differences intact up to a constant. That constant falls in the \ac{DC} bin of the \ac{DFT} that follows and is discarded by the demapper \cite{hernando2022frequency}.\\
The top panel shows the deviation region set by $m$, kept strictly inside the ambiguity boundary so every step is read correctly. In the bottom panel the region extends past that boundary, and a step exceeding $\pi$, marked at the lower left, is read as a smaller step of opposite sign and decoded on the opposite side of the circle.

\subsection{Communication Receiver}
\label{ssec:discriminator}
At the receiver, an FM demodulator recovers $x[n]$ and then applies conventional \ac{OFDM} processing \cite{hernando2022frequency}.
The received signal after down conversion is $z[n]=s[n]\,e^{j(2\pi\epsilon n+\psi)}$, where $\epsilon$ is a residual \ac{CFO} in cycles per sample and $\psi$ an unknown constant phase \cite{hernando2022frequency}. The discriminator forms
\begin{equation}
    \hat f[n] = \frac{1}{2\pi}\arg\big(z[n]\,z^*[n-1]\big)
    = f[n] + \epsilon ,
    \label{eq:discriminator}
\end{equation}
valid while $|f[n]+\epsilon|<\tfrac12$. The subcarrier demapper then discards $\epsilon$ without estimating it, since $k_0\ge1$ keeps the \ac{DC} bin out of $\mathcal{K}$, and phase noise with it. 

The product in \eqref{eq:discriminator} subtracts before taking the argument, so it returns $f[n]$ with no unwrapping stage. This holds on any sample that does not wrap, and fails with probability $P_w$ from \eqref{eq:wrap_prob}. The slow time estimator of Section~\ref{sec:receiver} applies the same product across symbols rather than samples, and inherits this property.

\subsection{Sensing Scenario}
\label{ssec:frame}
Sensing runs over a coherent processing interval of $U$ consecutive \ac{FM-OFDM} symbols, indexed $u=0,\dots,U-1$, each built from an independent draw of \ac{QAM} symbols $X_u[k]$. The data therefore changes from one symbol to the next. Symbols are separated by a guard of $N_g$ samples, so
\begin{equation}
    T_s = \frac{1}{f_s}, \qquad T_u = N T_s, \qquad
    T_{sym} = (N+N_g)\,T_s ,
    \label{eq:timing}
\end{equation}
with $f_s$ the sampling rate, $T_u$ the useful block duration and $T_{sym}$ the full symbol duration. The guard exceeds the largest scatterer delay, so consecutive symbols do not overlap at the receiver. All sensing processing works on the $N$ sample useful block, so the analysis does not depend on whether the guard is realized as zero padding \cite{hernando2025channel} or as a \ac{CP}.

The radar is monostatic, with the transmit and receive antennas separated enough for the self interference to be neglected \cite{chen2022antenna}, and the sensing receiver knows $s_u[n]$ exactly.\\
The receiver observes $L$ resolvable scatterers indexed $\ell=0,\dots,L-1$, where scatterer $\ell$ has complex amplitude $a_\ell$, round trip delay $\tau_\ell=p_\ell T_s$ with $p_\ell$ being as the corresponding delay bin, and radial velocity $v_\ell$ giving the Doppler shift
\begin{equation}
    \nu_\ell = \frac{2 v_\ell}{\lambda}, \qquad \lambda = \frac{c}{f_c} ,
    \label{eq:doppler}
\end{equation}
with $f_c$ the carrier frequency and $c$ the speed of light. Noise is $w[n]\sim\mathcal{CN}(0,\sigma_w^2)$, independent across samples and symbols, and the transmit amplitude is $A_c=\sqrt{P_t}$, so the instantaneous power is $P_t$ at every sample.3


\subsection{Bandwidth and Modulation Index}
\label{ssec:b99}

The transmit signal is a pure phase modulation, $s[n] = A_c e^{j\phi[n]}$, so its \ac{PSD} is the \ac{DTFT} of the autocorrelation $R[\Delta p] = \mathbb{E}\{s[n]s^*[n-\Delta p]\}$. $R[\Delta p]$ depends only on the phase difference, which is a Gaussian sum of $N_a/2$ symbols. Its characteristic function gives
\begin{equation}
    R[\Delta p] = A_c^2\,e^{-D_\phi(\Delta p)/2},\ D_\phi(\Delta p) \triangleq \mathrm{Var}\big(\phi[n]-\phi[n-\Delta p]\big).
    \label{eq:R_via_Dphi}
\end{equation}
The spectrum is therefore fixed by a single scalar function.\\ 
The phase $\phi[n] = 2\pi\sum_{i\le n}f[i]$ is a random walk with unbounded variance. Its lag difference is well behaved because every term older than $\Delta p$ appears at both endpoints and cancels. Using $f[n] = m\sqrt{N/N_a}\,x[n]$ from \eqref{eq:freq_cmd},
\begin{equation}
    \phi[n]-\phi[n-\Delta p] = 2\pi m\sqrt{\tfrac{N}{N_a}}\!\!\sum_{i=n-\Delta p+1}^{n}\!\! x[i],
    \label{eq:lag_diff}
\end{equation}

so $D_\phi$ is the variance of a length $\Delta p$ moving sum of the signal. Substituting the \ac{IDFT} \eqref{eq:idft} for $x[i]$ and swapping the two sums gives
\begin{equation}
    \sum_{i=n-\Delta p+1}^{n} x[i] = \sum_{k\in\mathcal{K}} X[k] \sum_{i=n-\Delta p+1}^{n} e^{j2\pi k i/N}.
    \label{eq:double_sum}
\end{equation}
Only the inner sum depends on the window. Setting $i = n-l$ pulls out $e^{j2\pi k n/N}$ and leaves a geometric series of $\Delta p$ terms with ratio $e^{-j2\pi k/N}$,
\begin{equation}
    G_{\Delta p}(k) = \sum_{l=0}^{\Delta p-1} e^{-j2\pi k l/N} = \frac{1-e^{-j2\pi k \Delta p/N}}{1-e^{-j2\pi k/N}}.
    \label{eq:geo_series}
\end{equation}

The term pulled out has magnitude one, so it does not change the variance. Subcarrier $k$ is therefore scaled by $G_{\Delta p}(k)$. Both terms of the fraction in \eqref{eq:geo_series} have the form $1-e^{-j\theta}$, and $|1-e^{-j\theta}|^2 = 4\sin^2(\theta/2)$ applies to each. The factors $4$ cancel and
\begin{equation}
    |G_{\Delta p}(k)|^2 = \frac{\sin^2(\omega_k \Delta p)}{\sin^2(\omega_k)}, \qquad \omega_k \triangleq \frac{\pi k}{N}.
    \label{eq:dphi_gain}
\end{equation}
The conjugate symmetry \eqref{eq:X_vec} makes each mirror pair contribute $2\,\mathrm{Re}\{X[k]G_{\Delta p}(k)\}$ with variance $\tfrac12|G_{\Delta p}(k)|^2$. The variances of the $N_a/2$ pairs add and
\begin{equation}
    D_\phi(\Delta p) = \frac{(2\pi m)^2}{N_a}\sum_{k\in\mathcal{K}}\frac{\sin^2(\omega_k \Delta p)}{\sin^2(\omega_k)}.
    \label{eq:dphi}
\end{equation}
At $\Delta p = 1$ every ratio in \eqref{eq:dphi} is unity and the sum returns $N_a$, so $D_\phi(1) = (2\pi\sigma_f)^2$, consistent with \eqref{eq:sigma_f}.
Equation \eqref{eq:dphi_gain} sets the spectrum shape. When $\omega_k \Delta p \ll 1$ both sines take their small argument value, the ratio goes to $\Delta p^2$, and the gain grows with the lag. The condition is on the product, so it holds for small $\Delta p$ and for small $\omega_k$. When $\omega_k \Delta p \gtrsim 1$ the numerator only swings between $0$ and $1$. The gain then stays below $1/\sin^2(\omega_k)$, and averages to $1/(2\sin^2\omega_k)$ over the lag. Growth ends at $\Delta p \sim 1/\omega_k$, which is one lag per subcarrier.

The active subcarriers occupy the lowest bins, so the highest one is $k \approx N_a/2$ and $\omega_{\max} \approx \pi N_a/(2N)$. It is the last gain to stop growing, and it does so at $\omega_{\max}\Delta p \sim 1$. The growth of $D_\phi$ therefore ends near
\begin{equation}
    \Delta p^\star \sim \frac{N}{N_a},
    \label{eq:crossover}
\end{equation}
this sets an order of magnitude, not an exact lag. This scale is also the correlation time of the signal, because a band of $N_a$ bins cannot change faster than one cycle per $N/N_a$ samples. Below $\Delta p^\star$ the phase builds up steadily and $D_\phi \simeq (2\pi m\Delta p)^2$. Above it the signal has been redrawn many times, the contributions are independent, and $D_\phi$ grows linearly. For $\Delta p$ below $\Delta p^\star$, \eqref{eq:R_via_Dphi} is a Gaussian in $\Delta p$ of width $1/(2\pi m)$, so the \ac{PSD} is Gaussian with standard deviation $m$, as \eqref{eq:sigma_f} requires. Its two-sided $99\%$ interval is
\begin{equation}
    \frac{B_{99}}{f_s} \approx 2\eta m = 5.15\,m, \qquad \eta \triangleq Q^{-1}(0.005) \approx 2.58 ,
    \label{eq:b99_law}
\end{equation}
a short lag limit of \eqref{eq:R_via_Dphi} and not a separate model. The absence of $N_a$ follows.
The dependence appears once $\Delta p^\star$ falls inside the decay length of $R$. The active bins are low, so $\omega_k$ is small and \eqref{eq:dphi_gain} expands in $\omega_k$ with $\Delta p$ kept exact, giving $|G_{\Delta p}(k)|^2 \simeq \Delta p^2[1-\tfrac13\omega_k^2(\Delta p^2-1)]$. Averaging over the active set replaces $\omega_k^2$ by $\overline{\omega^2} = (\pi/N)^2[k_0^2+k_0N_a/2+N_a^2/12]$. The $-1$ is smaller than the $\Delta p^2$ beside it and is dropped. Writing the lag as $y = 2\pi m\Delta p$ and expanding the exponential in \eqref{eq:R_via_Dphi} to first order leaves a single parameter,
\begin{equation}
    R(y) \simeq A_c^2\,e^{-y^2/2}\Big(1+\tfrac{1}{18}\zeta^2y^4\Big), \qquad \zeta^2 \triangleq \frac{3\,\overline{\omega^2}}{(2\pi m)^2},
    \label{eq:R_scaled}
\end{equation}
for $k_0 \ll N_a$, $\zeta = N_a/(4mN)$, the signal band edge divided by twice the \ac{RMS} frequency deviation. Only the first-order term is kept because $\zeta < 1$ over the range of interest.

The transform of \eqref{eq:R_scaled} needs the Gaussian transform pair and its fourth derivative, since $d^4\cos(\chi y)/d\chi^4 = y^4\cos(\chi y)$. Writing $\chi = F/m$ and taking $\mathcal{G}(\chi)$ as the standard normal density, the Rodrigues relation $\mathcal{G}^{(n)} = (-1)^n He_n\,\mathcal{G}$ \cite[Eq.~12.6.4]{cramer1999mathematical}, \cite[Eq.~22.11.8]{abramowitz1948handbook} turns the fourth derivative of $e^{-\chi^2/2}$ into the fourth probabilists' Hermite polynomial $He_4(\chi) = \chi^4 6\chi^2+3$ \cite[Eq.~12.6.5]{cramer1999mathematical}, so
\begin{equation}
    \Phi_s(\chi) \propto \mathcal{G}(\chi)\Big[1+\tfrac{1}{18}\zeta^2 He_4(\chi)\Big].
    \label{eq:gram_charlier}
\end{equation}
Equation \eqref{eq:gram_charlier} is the Gram-Charlier A series \cite[Eq.~17.6.5]{cramer1999mathematical} with excess kurtosis $\gamma_2 = \tfrac43\zeta^2$ from \cite[Eq.~17.6.6]{cramer1999mathematical}. The skewness is zero, so no lower order term appears. Orthogonality \cite[Eq.~12.6.6]{cramer1999mathematical} gives $\int\mathcal{G}He_4\,d\chi = 0$, so the \ac{RMS} bandwidth stays at $m$ for every $N_a$ and only the tails move. Placing $0.005$ of the power in each tail of \eqref{eq:gram_charlier} needs the tail integral $\int_{\chi_0}^{\infty}\mathcal{G}He_4\,d\chi = He_3(\chi_0)\mathcal{G}(\chi_0)$ with $He_3(\chi) = \chi^3-3\chi$. Writing $\chi_{99} = \eta+\delta$ and linearising $Q$ about $\eta$ cancels $\mathcal{G}(\eta)$ and leaves $\delta = \tfrac{1}{18}\zeta^2(\eta^3-3\eta)$, so with $C \triangleq (\eta^2-3)/18 = 0.2019$ and $B_{99}/f_s = 2m\chi_{99}$,
\begin{equation}
    \frac{B_{99}}{f_s} = 2\eta m\,\big(1+C\zeta^2\big).
    \label{eq:b99}
\end{equation}
Nothing in \eqref{eq:b99} is fitted, and it matches a direct numerical evaluation of \eqref{eq:R_via_Dphi} to better than $0.5\%$ for $\zeta \le 0.5$ and to $1.7\%$ at $\zeta = 1$.

Equation \eqref{eq:b99} can now be inverted for the modulation index. Setting $B_{99} = B_{ch}$, with $B_{ch}$ the occupied bandwidth budget of the assigned channel, and writing $m_0 = B_{ch}/(2\eta f_s)$ for the uncorrected value turns \eqref{eq:b99} into a quadratic in $m$, whose relevant root is
\begin{equation}
    m_{ch} = \frac{m_0}{2}\left[1+\sqrt{1-\frac{4C}{m_0^2}\Big(\frac{N_a}{4N}\Big)^{\!2}}\;\right].
    \label{eq:m_ch}
\end{equation}
Two further limits apply. Aliasing requires $B_{99} < f_s$, hence $m < 1/(2\eta)$. Reliable discrimination requires at most one phase wrap per symbol, which from \eqref{eq:wrap_prob} gives $m \le 1/[2Q^{-1}(1/2N)]$. The usable index is the smallest of the three,
\begin{equation}
    m_{\max} = \min\Big\{\tfrac{1}{2\eta},\;\big[2Q^{-1}(1/2N)\big]^{-1},\;m_{ch}\Big\}.
    \label{eq:m_max}
\end{equation}
For the parameters of Table~\ref{tab:params} these evaluate to $0.194$, $0.143$ and $0.0744$, so the bandwidth budget binds by a factor of two and neither aliasing nor wrapping is active. Equation~\eqref{eq:m_ch}  fixes the modulation index at $m = 0.0744$, used throughout.

\section{Range-Doppler Processing and the Ambiguity Function}
\label{sec:classical_failure}

In standard \ac{OFDM} radar processing, the received signal is the channel multiplied by the known data,
\begin{equation}
    R_u[k]=H_u[k]X_u[k]+W_u[k],
    \label{eq:factorization}
\end{equation}
where $R_u[k]$, $H_u[k]$ and $W_u[k]$ are the received signal, channel response and noise on subcarrier $k$ of symbol $u$ respectively. 
Dividing by $X_u[k]$ isolates $H_u[k]$, where the delay is a phase ramp across $k$ and the Doppler shift a phase progression across $u$, so a two-dimensional \ac{FFT} produces a range-Doppler map. For \ac{CP-OFDM} this division outperforms correlation processing, since removing the data removes its contribution to the correlation floor \cite{sturm2011waveform}.\\
Equation \eqref{eq:factorization} does not hold for the transmitted \ac{FM-OFDM} signal, since \eqref{eq:phase} maps $X[k]$ to $s[n]$ nonlinearly. Demodulation restores it, and ranging still works, but the Doppler shift falls in the \ac{DC} bin and the demapper throws it away, leaving the second \ac{FFT} nothing to transform. Matched filtering avoids this, at the cost of a data-dependent sidelobe floor characterized in Section~\ref{ssec:matched_filter}.

\subsection{Received Signal and Demodulation}
\label{ssec:classical_recipe}

Over one $N$-sample useful block, the $L$ scatterers of Section~\ref{ssec:frame} give the blockwise-circular, time-varying multipath channel
\begin{equation}
    h_u[n,p] \triangleq \sum_{\ell=0}^{L-1} a_\ell\, e^{j2\pi\nu_\ell (uT_{sym}+nT_s)}\, \delta[p-p_\ell] ,
    \label{eq:channel_model}
\end{equation}
so that the received signal in symbol $u$ is
\begin{equation}
    r_u[n] = \sum_{\ell=0}^{L-1} a_\ell\, e^{j2\pi\nu_\ell (uT_{sym}+nT_s)}\, s_u[n-p_\ell] + w_u[n] .
    \label{eq:received}
\end{equation}
The channel is assumed underspread over the useful block,
\begin{equation}
    f_D^{\max}\, T_u \ll 1 ,
    \label{eq:underspread}
\end{equation}
with $f_D^{\max}=\max_\ell|\nu_\ell|$, so that the intra-block Doppler phase is approximately constant, while the slow-time phase $e^{j2\pi\nu_\ell uT_{sym}}$ varies across symbols and is preserved for Doppler processing.

To apply the classical \ac{OFDM} radar processing chain to \ac{FM-OFDM}, the received waveform is first passed through the discriminator, which recovers the modulating signal $x[n]$. For a single scatterer, the discriminator output is
\begin{equation}
    g[n] = \frac{1}{2\pi}\arg\big(r[n]\,r^*[n-1]\big),
    \label{eq:disc_arg}
\end{equation}
which, after substituting \eqref{eq:received}, becomes
\begin{equation}
    g[n] = \nu_\ell T_s + m\sqrt{N/N_a}\,x[n-p_\ell] + \xi[n],
    \label{eq:disc_single}
\end{equation}
where $\xi[n]$ is zero-mean discriminator noise. The delay appears as a time shift of the modulating signal, while the Doppler shift appears as an additive constant.

Taking the \ac{DFT} of \eqref{eq:disc_single} over the useful block gives
\begin{equation}
    G[k] = \nu_\ell T_s N\,\delta[k] + m\sqrt{N/N_a}\,X[k]\,e^{-j2\pi k p_\ell/N} + \Xi[k],
    \label{eq:disc_dft}
\end{equation}
where $\Xi[k]\triangleq\mathrm{DFT}\{\xi[n]\}[k]$. The Doppler shift is confined to the constant term $\nu_\ell T_s N\,\delta[k]$, which is nonzero only at $k=0$. Since the active set $\mathcal{K}$ excludes $k=0$, that term is discarded, and the surviving subcarriers carry delay information but no velocity information.

After the \ac{DFT} and division by $X[k]$, the remaining phase term $H[k]=m\sqrt{N/N_a}\,e^{-j2\pi k p_\ell/N}$ is exactly the delay phase ramp, just like in classical \ac{CP-OFDM} radar. Taking the \ac{IDFT} over the active subcarriers $\mathcal{K}$ gives a range profile with a peak at $p_\ell$. However, the resolution is limited to $c/(2N_a\Delta f)$, not the full waveform bandwidth $c/(2B_{99})$ of \eqref{eq:range_resolution}. The loss factor is
\begin{equation}
    \frac{B_{99}}{N_a \Delta f} = \frac{B_{99}T_u}{N_a},
    \label{eq:res_penalty}
\end{equation}
here at $m=0.6/(2\pi)$, $N=512$ and $N_a=64$ this equals $3.9$. The demodulation step discards the wideband spectral spreading of the FM waveform because the delay is measured on the modulating signal $x[n]$, not on the transmitted FM signal $s[n]$.

For multiple targets, several echoes arrive together; the discriminator does not sum them linearly. Instead, the strongest echo dominates the rest, and weaker echoes are suppressed by more than their power difference. Combined with the Doppler null, only the strongest target appears, and it is confined to zero Doppler.\\
The direct application of the conventional \ac{CP-OFDM} 2D FFT chain, without the discriminator, also fails. In \ac{CP-OFDM}, the transmitted spectrum is flat, so dividing by $S_u[k]$ recovers $H_u[k]$ cleanly. For \ac{FM-OFDM} the spectrum is highly uneven and contains many deep nulls. Dividing by $S_u[k]$ gives a noise term $W_u[k]/S_u[k]$. At bins where $|S_u[k]|$ is near zero, this term becomes very large, and the range profile is noise only. Matched filtering, avoids both failures, and it correctly separates the two targets.

\subsection{Pulse Compression}
\label{ssec:matched_filter}

Demodulation loses Doppler, and direct spectral division amplifies noise, \ac{MF} avoids both problems by correlating the received signal with the known waveform. For symbol $u$, the aperiodic cross correlation is
\begin{equation}
    C_u[p] \triangleq \sum_{n} r_u[n]\, s_u^*[n-p],
    \label{eq:matched_filter}
\end{equation}
taken over the useful block. An \ac{FFT} based implementation needs zero-padding to $N'\ge2N$ to avoid circular correlation.
\begin{figure}
    \centering
    \includegraphics[width=1\linewidth]{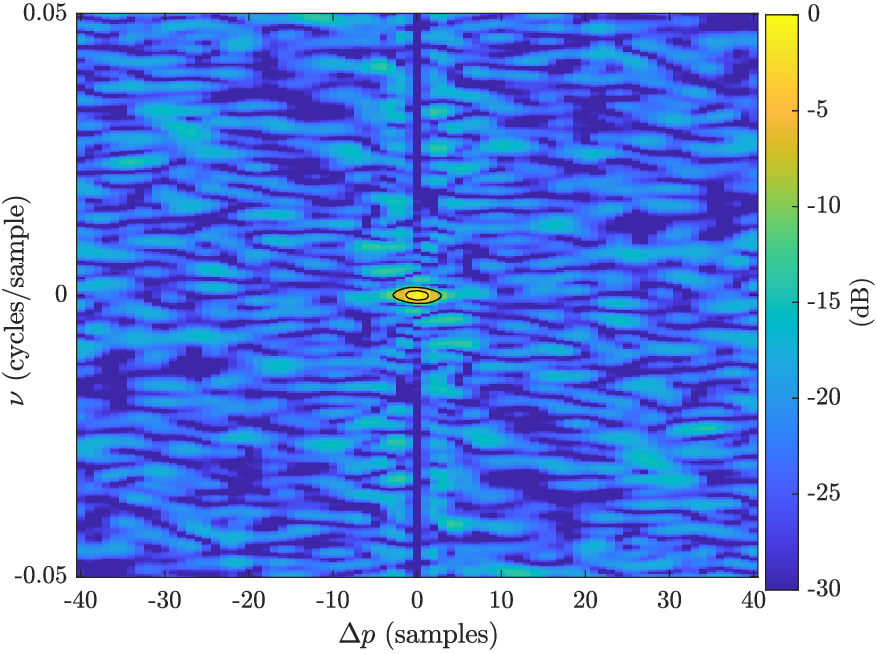}
    \caption{$|A(\Delta p,\nu)|$ of \eqref{eq:ambiguity}}
    \label{fig:af}
\end{figure}
Evaluating the \ac{MF} with one target gives its fundamental response before extending to multiple targets. Substituting \eqref{eq:received} with $L=1$ into \eqref{eq:matched_filter} gives
\begin{equation}
    C_u[p] = a_\ell\, e^{j2\pi\nu_\ell u T_{sym}}\, e^{j2\pi\nu_\ell p T_s}\, A(p_\ell-p,\nu_\ell) + \tilde w_u[p],
    \label{eq:mf_output}
\end{equation}
where
\begin{equation}
    A(\Delta p,\nu) = \sum_{n} s^*[n]\, s[n-\Delta p]\, e^{j2\pi\nu n T_s}
    \label{eq:ambiguity}
\end{equation}
is the \ac{AF}. The phase factor $e^{j2\pi\nu_\ell p T_s}$ is constant for a fixed range bin and is absorbed into the slow-time amplitude. The output of pulse compression is therefore completely described by $A(\Delta p,\nu)$.

The zero-Doppler cut $A(\Delta p,0)$ is the standard autocorrelation used for ranging, but it only describes the response at zero Doppler. The matched filter output is evaluated at the target Doppler, so the full two-dimensional ambiguity function is needed. Since $|s[n]|=A_c$ for every sample, the data can only enter the ambiguity function through the phase increment,
\begin{equation}
\begin{split}
    A(\Delta p,\nu) &= A_c^2 \sum_{n} e^{-j\Delta_{\Delta p}\phi[n]}\, e^{j2\pi\nu n T_s},\\
    \Delta_{\Delta p}\phi[n] &= \phi[n]-\phi[n-\Delta p] = 2\pi\!\!\!\!\!\!\!\!\sum_{i=n-\Delta p+1}^{n}\!\!\!\!\!\! f[i].
\end{split}
    \label{eq:af_phase_form}
\end{equation}
The phase change over a lag of $\Delta p$ samples determines the ambiguity function, and its two cuts behave differently. Fig.~\ref{fig:af} shows $|A(\Delta p,\nu)|$ near the origin for $k_0=1$ and $m=0.6/(2\pi)$. The vertical band at $\Delta p=0$ is deterministic and identical for every data realization, while the surrounding sidelobe field changes with the data.

\subsubsection{Zero-Delay Cut}

At $\Delta p=0$, the phase increment in \eqref{eq:af_phase_form} is zero for every sample. The ambiguity function reduces to
\begin{equation}
    A(0,\nu) = A_c^2\sum_{n=0}^{N-1} e^{j2\pi\nu n T_s}
    = A_c^2\, e^{j\pi\nu(N-1)T_s}\,
    \frac{\sin(\pi\nu N T_s)}{\sin(\pi\nu T_s)} ,
    \label{eq:mainlobe}
\end{equation}
this is the Dirichlet kernel of order $N$. Its peak is $A(0,0)=A_c^2N$, and its Doppler mainlobe width and sidelobe structure are fixed for every data realization because $|s[n]|=A_c$ for all $n$. The zero-delay cut therefore does not depend on the data.

For CP-OFDM, the zero-delay cut changes from block to block because it depends on the transmitted data. The results in \cite{liu2025cp} are averaged over many realizations. Here, \eqref{eq:mainlobe} is exact for every block. Simulation confirms the difference. Across independent realizations, the peak-to-peak spread of $A(0,\nu)$ is of order $10^{-16}$ for FM-OFDM, while the same measurement for CP-OFDM gives a spread of order $10^{-1}$. The optimality result in \cite{liu2025cp} applies to linearly modulated waveforms, and FM-OFDM is not in that class because the data are mapped to the signal nonlinearly. Section~\ref{sec:results} runs a bandwidth-matched comparison.

\subsubsection{Nonzero-Delay Cut}
Away from $\Delta p=0$ , the phase increment depends on the data. Applying \eqref{eq:R_via_Dphi} to \eqref{eq:af_phase_form} gives the mean phase term $e^{-D_\phi(\Delta p)/2}$, with $D_\phi$ from \eqref{eq:dphi}, so the mean ambiguity function is
\begin{equation}
    \mathbb{E}\big[A(\Delta p,\nu)\big]
    = A_c^2\, e^{-D_\phi(\Delta p)/2} \sum_{n} e^{j2\pi\nu n T_s} .
    \label{eq:mean_af}
\end{equation}

The variance $D_\phi(\Delta p)$ grows with $\Delta p$, but it does not grow without bound. As $\Delta p$ grows, the numerator $\sin^2(\pi k\Delta p/N)$ oscillates between $0$ and $1$. Its average is $1/2$, so for large $\Delta p$ the variance approaches
\begin{equation}
\begin{split}
    \bar D_\phi &= \frac{(2\pi m)^2}{2N_a} \sum_{k\in\mathcal{K}}
    \frac{1}{\sin^2(\pi k/N)}\\
    &\approx \frac{4 m^2 N^2}{N_a}
    \sum_{k=k_0}^{k_0+N_a/2-1} \frac{1}{k^2} ,
\end{split}
    \label{eq:dphi_plateau}
\end{equation}
where the approximation $\sin(\pi k/N)\approx\pi k/N$ holds for $k_0+N_a/2 \ll N$. The phase increment variance approaches this floor only after the lag is large enough for the slowest active subcarrier, located at $k_0$, to complete several cycles. The residual oscillation therefore decays on a lag scale of roughly $N/k_0$.

\begin{figure}
    \centering
    \includegraphics[width=1\linewidth]{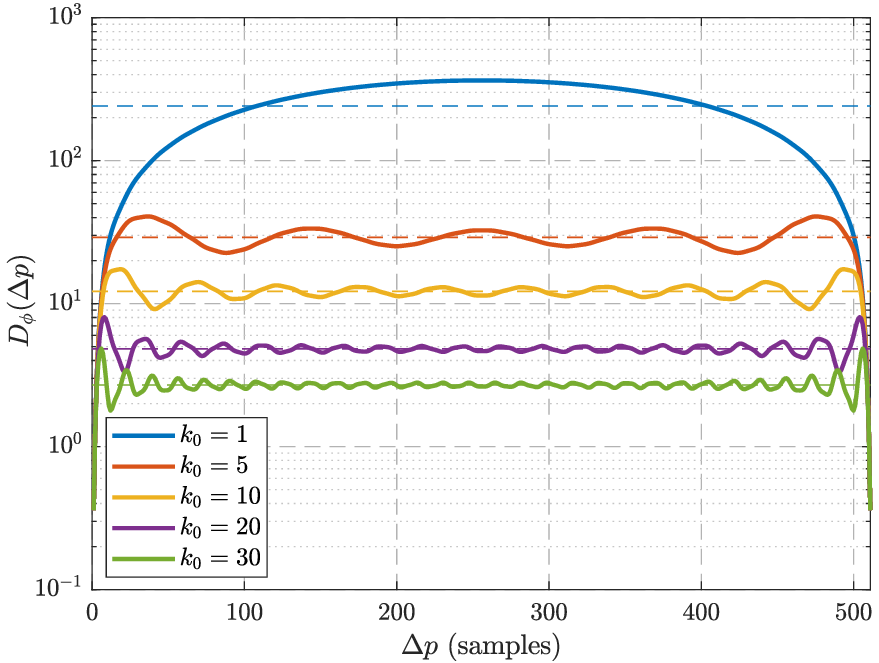}
    \caption{Phase increment variance $D_\phi(\Delta p)$ for several $k_0$.}
    \label{fig:dphi}
\end{figure}

The gain $|G_{\Delta p}(k)|^2$ of \eqref{eq:dphi_gain} is largest at small $k$, where the denominator $\sin(\pi k/N)$ is small, so low-frequency subcarriers contribute the most to $D_\phi(\Delta p)$. At $k=0$ the ratio in \eqref{eq:dphi} tends to $\Delta p$, so a \ac{DC} subcarrier would contribute $(2\pi m)^2\Delta p^2/N_a$ and never saturate. Every active term is instead capped at $1/\sin^2(\pi k/N)$, so $k_0\ge1$ is what sets the floor. Fig.~\ref{fig:dphi} shows $D_\phi(\Delta p)$ for several values of $k_0$, with the floor level $\bar D_\phi$ from \eqref{eq:dphi_plateau} shown as dashed lines.

The mean ambiguity function gives only the coherent part of the sidelobe response. The actual sidelobe level is set by the second moment. Normalizing to the mainlobe peak gives the mean-square zero-Doppler sidelobe level
\begin{equation}
\begin{split}
    \mathrm{SLL}(\Delta p) &\triangleq
    \frac{\mathbb{E}\big[|A(\Delta p,0)|^2\big]}{|A(0,0)|^2}\\
    &\approx \rho^2\, e^{-D_\phi(\Delta p)}
    + \rho\,\frac{1-e^{-D_\phi(\Delta p)}}{B_{99}T_u} ,
\end{split}
    \label{eq:sll}
\end{equation}
where $\rho \triangleq 1-|\Delta p|/N$ is the overlap fraction of the aperiodic correlation. The first term comes from the mean part of the phase increment, which adds coherently over the overlapping samples.The second term comes from the fluctuations and
sets the incoherent floor. Its level is fixed by the equivalent noise bandwidth
\begin{equation}
    B_e \triangleq \frac{\big(\int \Phi_s(f)\,df\big)^2}{\int \Phi_s^2(f)\,df}
    = 2\sqrt{\pi}\,m f_s ,
    \label{eq:be}
\end{equation}
which counts the independent spectral degrees of freedom of the signal. The closed form follows from the Gaussian spectrum of \eqref{eq:gram_charlier}, and gives $B_e/B_{99} = 2\sqrt{\pi}/(2\eta) = 0.688$, so the floor sits $1.6$~dB above what the $99\%$ bandwidth alone would suggest. For small $|\Delta p|$, $\rho\to1$, and using the saturation level from \eqref{eq:dphi_plateau}, the two limits are
\begin{equation}
    \mathrm{SLL} \approx
    \begin{cases}
        -10\log_{10}\big(B_e T_u\big)\ \mathrm{dB}, &
        \bar D_\phi \gg 1,\\[2pt]
        -4.34\,\bar D_\phi\ \mathrm{dB}, & \bar D_\phi \lesssim 1 .
    \end{cases}
    \label{eq:sll_m}
\end{equation}
Substituting \eqref{eq:be} writes the first branch as $-10\log_{10}(2\sqrt{\pi}mN)$, so in the saturated regime the far sidelobe floor is fixed by the modulation index and the block length alone. The exact second moment follows because the phase increments are jointly Gaussian under the same approximation. Their covariance is
\begin{equation}
    C_{\Delta p}(\tau) = \frac{(2\pi m)^2}{N_a}
    \sum_{k\in\mathcal{K}}
    \frac{\sin^2(\pi k\Delta p/N)}{\sin^2(\pi k/N)}
    \cos\!\Big(\frac{2\pi k\tau}{N}\Big) ,
    \label{eq:cov_tau}
\end{equation}
with $C_{\Delta p}(0)=D_\phi(\Delta p)$. The double sum then gives
\begin{equation}
    \frac{\mathbb{E}\big[|A(\Delta p,0)|^2\big]}{|A(0,0)|^2}
    = \frac{1}{N^2}\sum_{|\tau|<\rho N}
    \big(\rho N-|\tau|\big)\,
    e^{-D_\phi(\Delta p)+C_{\Delta p}(\tau)} .
    \label{eq:sll_exact}
\end{equation}

\begin{figure}
    \centering
    \includegraphics[width=1\linewidth]{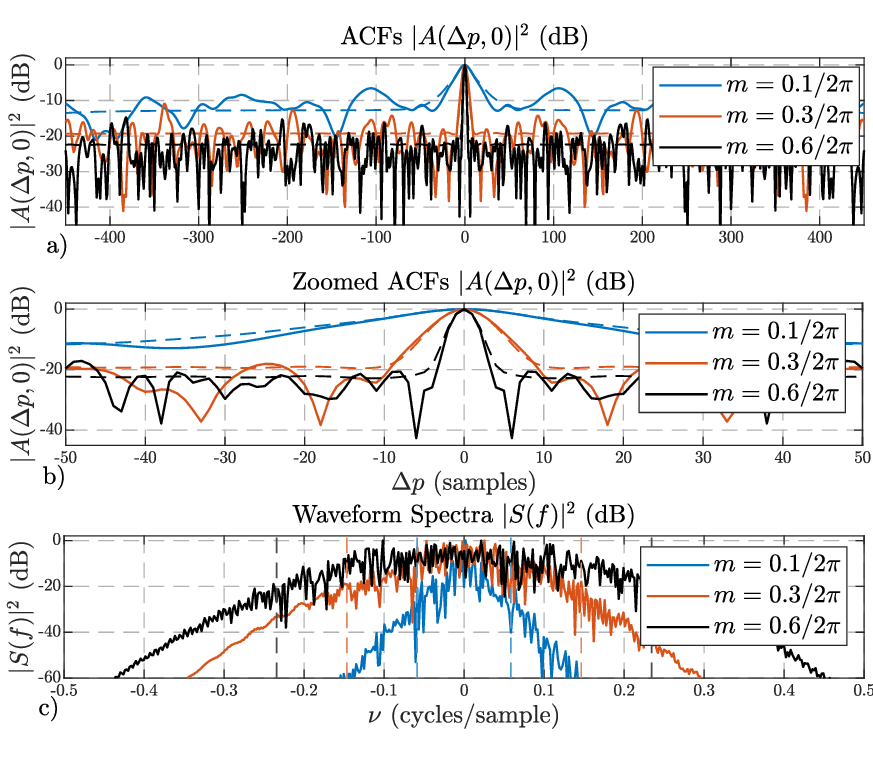}
    \caption{Zero Doppler cut and sidelobe floor for three modulation indices.}
    \label{fig:acf_m}
\end{figure}

Simulation confirms \eqref{eq:sll_exact} within $0.5$~dB across $m\in[0.1,0.6]/(2\pi)$. The approximate form \eqref{eq:sll} captures the scaling but not the exact constant. Because $D_\phi(\Delta p)$ starts from zero and then saturates, the coherent term is elevated just outside the mainlobe. This produces a skirt that extends until $D_\phi(\Delta p)$ passes $\ln(B_e T_u)$, at the operating point that is $\Delta p \approx 5$ samples. Closely spaced scatterers are resolved against that skirt rather than against the far-lag floor. Fig.~\ref{fig:acf_m} shows this for different modulation indices.

Raising $k_0$ also degrades the \ac{MF} sidelobe floor. As $k_0$ increases, the term at the lowest active subcarrier is replaced by a term at a higher subcarrier. Since the denominator in \eqref{eq:dphi_plateau} increases with $k$, the removed term is larger, so $\bar D_\phi$ decreases. The coherent sidelobe term $e^{-\bar D_\phi}$ therefore increases with $k_0$.

Raising $k_0$ improves link robustness \cite{hernando2022frequency} at two costs in sensing. It removes the Doppler observable from the demodulated path, and it reduces $\bar D_\phi$, which raises the coherent sidelobe term. Section~\ref{ssec:sim_sll} shows that this term stays far below the incoherent floor of \eqref{eq:sll_m} across the admissible range, so $k_0$ is free there.

Data dependent sidelobes are not unique to \ac{FM-OFDM}. In \ac{CE-OFDM}, the autocorrelation sidelobes also vary with the transmitted symbols \cite{11130717}. The zero-delay mainlobe result here is deterministic and holds for every realization, unlike the statistical mainlobe model in \cite{11130717}, which includes data-dependent range-Doppler coupling \cite{thompson2009constant}.

\subsection{Differential Slow-Time Processing}
\label{ssec:doppler_consequence}

At a detected range bin $\hat\tau_\ell$, the pulse-compressed output across symbols is
\begin{equation}
\begin{split}
     y_\ell[u] &\triangleq C_u[\hat p_\ell]\\
    &\approx \alpha_\ell\, e^{j(2\pi\nu_\ell u T_{sym} + \theta_\ell[u])}
    + w_\ell[u] ,
\end{split}
    \label{eq:slow_time_sample_preview}
\end{equation}
where $\theta_\ell[u]$ collects the off-grid and leakage contributions. Its variance follows from \eqref{eq:sll} and is quantified in Section~\ref{ssec:slow_time_samples}. Because the data symbols $X_u[k]$ change with every symbol, $\theta_\ell[u]$ varies pseudo-randomly across $u$ and cannot be removed by a constant phase correction.

Doppler information is restored by the \ac{MF}, but the slow-time sample still contains this data-dependent phase. Section~\ref{ssec:differential_doppler} estimates the slow-time phase increments instead of the phase itself, this confines $\theta_\ell[u]$ to the first differences of the estimator weights and needs no knowledge of the data and no phase unwrapping.

\section{Proposed Differential Sensing Receiver}
\label{sec:receiver}

Section~\ref{sec:classical_failure} showed that the \ac{MF} preserves range while the slow-time phase carries data-dependent fluctuations. The proposed receiver keeps the \ac{MF} for range and estimates Doppler from the slow-time phase progression. That progression survives because the \ac{MF} correlates the received samples against the known transmitted waveform and never applies the discriminator, so the Doppler is never collapsed onto a subcarrier index. The estimator reuses the product of \eqref{eq:disc_arg}. The index is the symbol $u$. It does not cancel the data-dependent phase. It confines that phase to the first differences of the weights, so one weight design controls both the noise and the data.

\subsection{Slow-Time Samples}
\label{ssec:slow_time_samples}

For each symbol $u$ the received block is matched-filtered against the known transmitted waveform, as in \eqref{eq:matched_filter}. Noncoherent averaging
\begin{equation}
    \bar C[p] = \frac{1}{U}\sum_{u=0}^{U-1} \big|C_u[p]\big|,
    \label{eq:noncoherent}
\end{equation}
reduces the scatter of the data-dependent sidelobes but not their mean level. Detection therefore sits on the floor of \eqref{eq:sll}. The detected peaks give the delay bins $\hat p_\ell$ at the resolution
\begin{equation}
    \Delta R = \frac{c}{2B_{99}},
    \label{eq:range_resolution}
\end{equation}
with $B_{99}$ from \eqref{eq:b99}.

At a detected bin, the pulse-compressed output across symbols is
\begin{equation}
    y_\ell[u] \triangleq C_u[\hat p_\ell]
    \approx \alpha_\ell\, e^{j(2\pi\nu_\ell u T_{sym} + \theta_\ell[u])} + w_\ell[u],
    \label{eq:slow_time_sample}
\end{equation}
where $w_\ell[u]$ is post-compression noise and $\theta_\ell[u]$ is the residual data-dependent phase at that bin.

Leakage from other scatterers dominates $\theta_\ell[u]$. Each interferer contributes its power ratio times the sidelobe level at its delay separation, halved because only the quadrature component moves the phase,
\begin{equation}
    \sigma_\theta^2 \approx \sigma_{\mathrm{og}}^2
    + \frac{1}{2}\sum_{q\neq\ell}\frac{|a_q|^2}{|a_\ell|^2}\,\mathrm{SLL}(\hat p_\ell - \hat p_q).
    \label{eq:sigma_theta}
\end{equation}
The term $\sigma_{\mathrm{og}}^2$ comes from evaluating the \ac{AF} at a fractional lag, when the detected peak does not fall on the true delay. It is largest at half a bin and vanishes on the grid. Equation~\eqref{eq:dphi} holds only at integer lag, so $\sigma_{\mathrm{og}}^2$ has no closed form and is taken from simulation. At the operating point it is $1.87\times10^{-6}$, against $3.11\times10^{-4}$ for a single interferer $10$~dB down at $\Delta p = 91$, where $\mathrm{SLL} = -22.48$~dB. Equation \eqref{eq:sigma_theta} matches simulation to within $0.42$~dB over two decades of interferer amplitude. The measured autocorrelation of $\theta_\ell[u]$ stays below $0.024$ in magnitude at every nonzero lag, so $\theta_\ell[u]$ is zero mean and white across $u$.

The model assumes one dominant scatterer per resolved bin, separated from the others beyond the sidelobe skirt. That skirt extends to $\Delta p \approx 6$ samples at the operating point. Two comparable scatterers in one bin make $y_\ell[u]$ the phase of a vector sum, and \eqref{eq:slow_time_sample} does not describe. Inside the skirt the leakage keeps a coherent component at mainlobe level, rotating at the Doppler difference between the two scatterers. That component is not white, and \eqref{eq:sigma_theta} departs by several dB. 

\subsection{Doppler Estimator}
\label{ssec:differential_doppler}
The unknown phase $\theta_\ell[u]$ changes with every symbol, so the absolute phase of $y_\ell[u]$ is unusable. Doppler appears as the rotation from one symbol to the next, and then the estimator measures by differencing consecutive symbols.

The phase change between two symbols is
\begin{equation}
    \Delta\varphi_\ell[u] \triangleq \arg\big(y_\ell[u]\, y_\ell^*[u-1]\big) \in (-\pi,\pi],
    \label{eq:increment}
\end{equation}
since multiplying by a conjugate subtracts phases. With \eqref{eq:slow_time_sample} this becomes
\begin{equation}
    \Delta\varphi_\ell[u] = 2\pi\nu_\ell T_{sym}
    + \big(\theta_\ell[u]-\theta_\ell[u-1]\big)
    + \big(\delta_\ell[u]-\delta_\ell[u-1]\big),
    \label{eq:increment_model}
\end{equation}
where $\delta_\ell[u]$ is the phase error from noise. Only the noise across $y_\ell[u]$ shifts its phase, and that is half the noise power, so $\delta_\ell[u]$ has variance $1/(2\gamma_\ell)$, where $\gamma_\ell = |\alpha_\ell|^2/\sigma_{w_\ell}^2$ is the post-compression \ac{SNR} at the detected bin.

The wanted term no longer depends on $u$. Every increment carries the same $2\pi\nu_\ell T_{sym}$, so averaging the increments is enough. Averaging with weights $w_u$ gives
\begin{equation}
    \hat\nu_\ell = \frac{1}{2\pi T_{sym}}
    \sum_{u=1}^{U-1} w_u\, \Delta\varphi_\ell[u],
    \label{eq:differential_estimator}
\end{equation}
and the radial velocity estimate is $\hat v_\ell = \lambda\hat\nu_\ell/2$. Each increment carries one copy of the wanted term, so the weights sum to one, $\sum_{u=1}^{U-1} w_u = 1$. The boundary convention is $w_0 = w_U = 0$.

Only differences appear in \eqref{eq:differential_estimator}, so no unwrapping is needed. Each difference must instead stay within one turn,
\begin{equation}
    \big|2\pi\nu_\ell T_{sym} + \theta_\ell[u]-\theta_\ell[u-1]
    + \delta_\ell[u]\big| < \pi \quad \forall u,
    \label{eq:nowrap}
\end{equation}
which is the condition for \eqref{eq:increment_model} to hold. Regrouping the sum in
\eqref{eq:differential_estimator} by symbol rather than by difference gives
\begin{equation}
    \sum_{u=1}^{U-1} w_u\big(\theta_\ell[u]-\theta_\ell[u-1]\big)
    = \sum_{u=0}^{U-1} \big(w_u - w_{u+1}\big)\,\theta_\ell[u].
    \label{eq:telescope}
\end{equation}
The data phase does not cancel. It is multiplied by the gap between neighbouring weights instead of by the weight itself, and the noise terms regroup the same way, so one set of gaps controls both.

The error is a weighted sum of the two perturbations,
\begin{equation}
    \hat\nu_\ell - \nu_\ell = \frac{1}{2\pi T_{sym}}
    \sum_{u=0}^{U-1}\big(w_u - w_{u+1}\big)
    \big(\theta_\ell[u] + \delta_\ell[u]\big).
    \label{eq:error}
\end{equation}
Both have zero mean, so the estimator is unbiased, $\mathbb{E}[\hat\nu_\ell]=\nu_\ell$. Both are white across $u$, so their variances add without cross terms. Writing
\begin{equation}
    S(w) \triangleq \sum_{u=0}^{U-1}\big(w_u - w_{u+1}\big)^2
    \label{eq:sw}
\end{equation}
gives
\begin{equation}
    \mathrm{Var}(\hat\nu_\ell)
    = \frac{\sigma_\theta^2 + 1/(2\gamma_\ell)}
    {(2\pi T_{sym})^2}\; S(w),
    \label{eq:estimator_variance}
\end{equation}
with $\sigma_\theta^2$ from \eqref{eq:sigma_theta}. Noise and data reach the estimate through the same $S(w)$, so one minimization serves both.

The weights that minimize \eqref{eq:sw} change slowly, vanish at both ends, and sum to one. These conditions give the parabolic window
\begin{equation}
    w_u^{\star} = \frac{6\,u\,(U-u)}{U(U^2-1)}, \qquad u = 1,\dots,U-1,
    \label{eq:optimal_weights}
\end{equation}
with $S(w^{\star}) = 12/[U(U^2-1)]$, so
\begin{equation}
    \mathrm{Var}(\hat\nu_\ell)\big|_{w^{\star}}
    = \frac{12\,\sigma_\theta^2 + 6/\gamma_\ell}
    {(2\pi T_{sym})^2\, U(U^2-1)}.
    \label{eq:optimal_variance}
\end{equation}
These are the weights Kay derived for a pure tone in additive noise \cite{kay1989fast}. They stay optimal here because $\theta_\ell[u]$ is white, and makes \eqref{eq:optimal_variance} a design equation.

The second term of \eqref{eq:optimal_variance} falls with transmit power and matches the \ac{CRB} for frequency estimation from $U$ coherent samples,
\begin{equation}
    \mathrm{CRB}(\nu_\ell)
    = \frac{6}{(2\pi T_{sym})^2\,\gamma_\ell\, U(U^2-1)},
    \label{eq:crb}
\end{equation}
so the estimator is efficient against thermal noise. The first term has no $\gamma_\ell$ and takes over at
\begin{equation}
    \gamma_\ell^{\mathrm{fl}} = \frac{1}{2\sigma_\theta^2},
    \label{eq:floor_crossover}
\end{equation}
beyond which the variance saturates at
\begin{equation}
    \mathrm{Var}(\hat\nu_\ell)\big|_{\mathrm{fl}}
    = \frac{12\,\sigma_\theta^2}{(2\pi T_{sym})^2\, U(U^2-1)}.
    \label{eq:floor_level}
\end{equation}
This is the sensing floor in the Doppler domain. It keeps the $U^{-3}$ decay of the noise term, so the floor falls as $U$ increases.

Uniform weights $w_u = 1/(U-1)$ recover the classical unweighted phase-difference estimator \cite{tretter1985estimating, kay1989fast}. Every interior gap in \eqref{eq:telescope} then vanishes and only the first and last symbols survive. That case gives $S = 2/(U-1)^2$ and a penalty of
\begin{equation}
    \frac{S(w_{\mathrm{unif}})}{S(w^{\star})}
    = \frac{U(U+1)}{6(U-1)} \;\approx\; \frac{U}{6},
    \label{eq:weight_penalty}
\end{equation}
is $10.42$~dB in variance at $U = 64$. Removing the interior data phases therefore costs more than it saves. In \ac{RMSE}, \eqref{eq:weight_penalty} makes the unweighted estimator decay as $U^{-1}$ and the weighted estimator as $U^{-3/2}$.

Raising $k_0$ leaves $\sigma_\theta^2$ unchanged over the admissible range. It raises the coherent sidelobe term of Section~\ref{ssec:matched_filter}, but that term stays far below the incoherent floor, so the level entering \eqref{eq:sigma_theta} does not move and neither does the floor \eqref{eq:floor_level}. The cutoff subcarrier therefore costs the demodulated Doppler observable and nothing in velocity accuracy.

\begin{figure}[!t]
    \centering
    \includegraphics[width=0.9\linewidth]{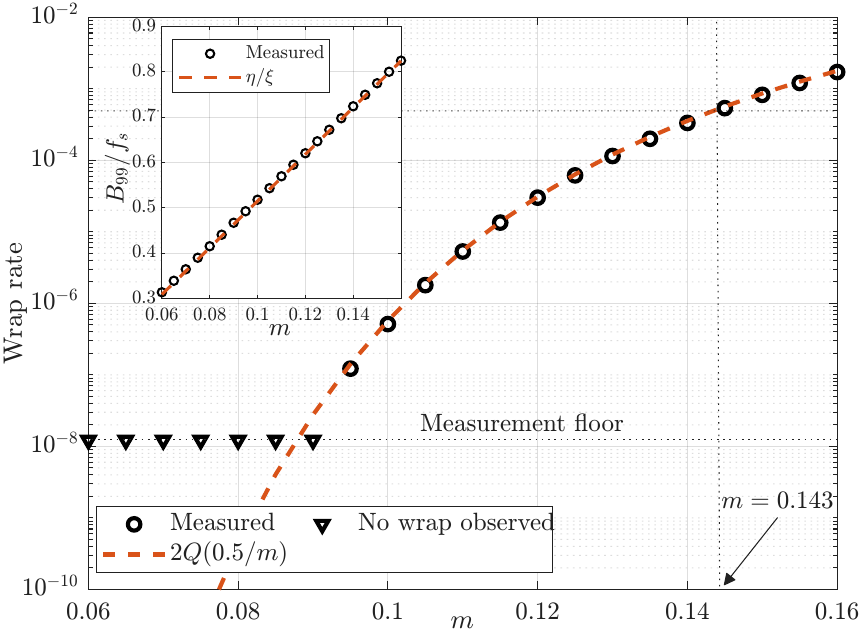}
    \caption{Wrap rate against $m$. Inset: occupancy against $\eta/\xi$.}
    \label{fig:mocc}
\end{figure}
\subsection{The Doppler Floor}
\label{ssec:doppler_floor}
We compared \eqref{eq:differential_estimator} under both weight choices against periodogram. The scene has a reference scatterer half a sample off the delay grid at $100$~m/s and one interferer $10$~dB down, $91$ bins away, at $-60$~m/s. The Post-compression \ac{SNR} is $18$~dB below integrated \ac{SNR} at $U=64$.

Above $\gamma_\ell = 12$~dB both estimators track \eqref{eq:crb} and agree to within $0.2$~dB, so the proposed estimator gains no accuracy. Both saturate at $0.0401$~m/s against $0.0384$~m/s from \eqref{eq:optimal_variance}. Two unrelated estimators stopping at the same value places the floor in the waveform. The variance doubles at $32.4$~dB, matching the $32.1$~dB of \eqref{eq:floor_crossover}. Uniform weights lose $10.46$~dB in variance, against $10.42$~dB from \eqref{eq:weight_penalty}.

Below $\gamma_\ell = 10$~dB the differential estimator loses. Differencing doubles the noise phase variance in each increment, so \eqref{eq:nowrap} fails while the periodogram peak still holds. The periodogram tracks \eqref{eq:crb} down to $2$~dB. Below $5$~dB uniform weights beat parabolic weights, since a wrapped increment near the block centre carries a larger weight in \eqref{eq:optimal_weights}. Below $1$~dB neither estimate is usable.

The value of \eqref{eq:differential_estimator} is not accuracy. It is that \eqref{eq:optimal_variance} predicts the floor and the crossover to within $0.5$~dB, in closed form through $\sigma_\theta^2$. The estimator is also search free and runs in $\mathcal{O}(U)$ operations.

The same condition \eqref{eq:nowrap} sets the unambiguous velocity,
\begin{equation}
    |\nu_\ell| < \frac{1}{2T_{sym}},
    \qquad |v_\ell| < \frac{\lambda}{4T_{sym}},
    \label{eq:ambiguity_limit}
\end{equation}
A coherent slow-time transform has the same limit, so differencing costs no coverage. A wrapped increment shifts $\hat\nu_\ell$ by $w_u/T_{sym}$, largest at the block centre.

The closed forms also need \eqref{eq:underspread}, since they assume block-constant path weights. Taking $f_D^{\max}T_u \le 0.1$ against $f_D^{\max}T_{sym} < 1/2$ leaves a factor of five between them. The estimator still works in that band, but intra-symbol Doppler rotation degrades the matched filter and inflates \eqref{eq:estimator_variance}. Section~\ref{sec:results} sweeps the band and reports $f_D^{\max}T_{sym}$ for every scenario.

\section{Simulation Results}
\label{sec:results}
\begin{figure}[!t]
    \centering
    \includegraphics[width=0.9\linewidth]{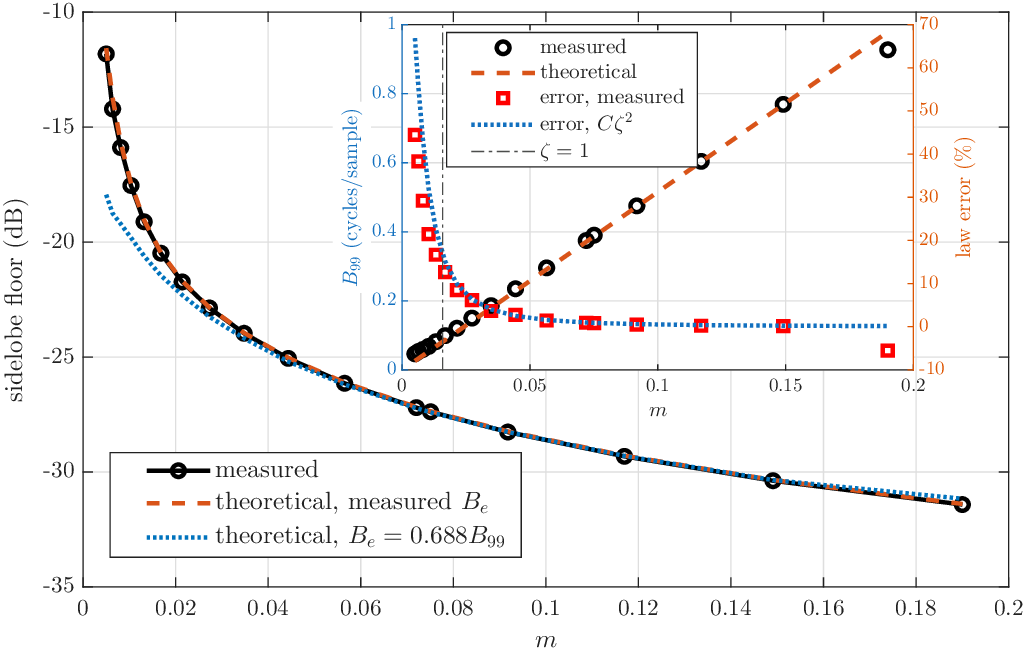}
    \caption{Sidelobe floor against $m$. Inset: occupied bandwidth against \eqref{eq:b99_law} and the error predicted by \eqref{eq:b99}.}
    \label{fig:b99}
\end{figure}
\begin{table}[b]
\caption{Simulation parameters}
\label{tab:params}
\centering
\begin{tabular}{llr}
\hline
\multicolumn{3}{l}{\emph{Grid}}\\
Carrier frequency & $f_c$ & $28$~GHz \\
Sample rate & $f_s$ & $245.76$~MHz \\
Subcarrier spacing & $\Delta f$ & $120$~kHz \\
Block length & $N$ & $2048$ \\
Guard length & $N_g$ & $0$ \\
Symbols per interval & $U$ & $112$ \\
Symbol duration & $T_{sym}$ & $8.33~\mu$s \\
Channel bandwidth & $B_{ch}$ & $95.04$~MHz \\
FM-OFDM active subcarriers & $N_a$ & $128$ \\
Cutoff subcarrier & $k_0$ & $1$ \\
Constellation & $M$ & $16$ \\
\hline
\multicolumn{3}{l}{\emph{Derived}}\\
FM-OFDM index, \eqref{eq:m_ch} & $m$ & $0.0744$ \\
Range resolution, \eqref{eq:range_resolution} & $\Delta R$ & $1.58$~m \\
Unambiguous range & & $1250$~m \\
Unambiguous velocity, \eqref{eq:ambiguity_limit} & & $321$~m/s \\
\hline
\end{tabular}
\end{table}

This section checks the closed forms of Sections~\ref{sec:Syst_mod} to \ref{sec:receiver} against simulation, and compares \ac{FM-OFDM} with \ac{CP-OFDM} and \ac{CE-OFDM} at equal occupied bandwidth and equal transmit power. The spectrum is checked first, since every later comparison rests on the bandwidth match. Table~\ref{tab:params} lists the parameters. The grid is a single \ac{FR2} channel of $66$ resource blocks that all three waveforms fill. Equation \eqref{eq:m_ch} fixes the \ac{FM-OFDM} index. The \ac{CP-OFDM} active subcarrier count and the \ac{CE-OFDM} index have no closed form and are iteratively adjusted until all three waveforms measure the same $B_{99}$.

Scatterer delays are applied as circular shifts of the transmitted block, so the model assumes a prefix longer than the largest delay but charges no time or energy for it, giving $N_g = 0$ and $T_{sym} = T_u$. The unambiguous range in Table~\ref{tab:params} is therefore the wrap limit of the circular model rather than the prefix limit of a deployed system. \ac{SNR} is quoted per sample at the receiver input, with compression adding $10\log_{10}N$ and integration over the interval a further $10\log_{10}U$. Each figure states its scatterers, and every scene holds $f_D^{\max}T_{sym}\le 0.1$.

\subsection{Spectrum and Occupied Bandwidth}
\label{ssec:sim_spectrum}
Fig.~\ref{fig:mocc} measures the wrap rate against $m$. The measurement follows $2Q(0.5/m)$ of \eqref{eq:wrap_prob} wherever the prediction is above the floor resolved by the simulated symbols, and no wrap is observed below it. One wrap per block is a rate of $1/N$, and the measured crossing matches the middle bound of \eqref{eq:m_max}. The operating point sits orders of magnitude below the floor, so phase wrapping plays no part in anything that follows.
\begin{figure}[!t]
    \centering
    \includegraphics[width=0.76\linewidth]{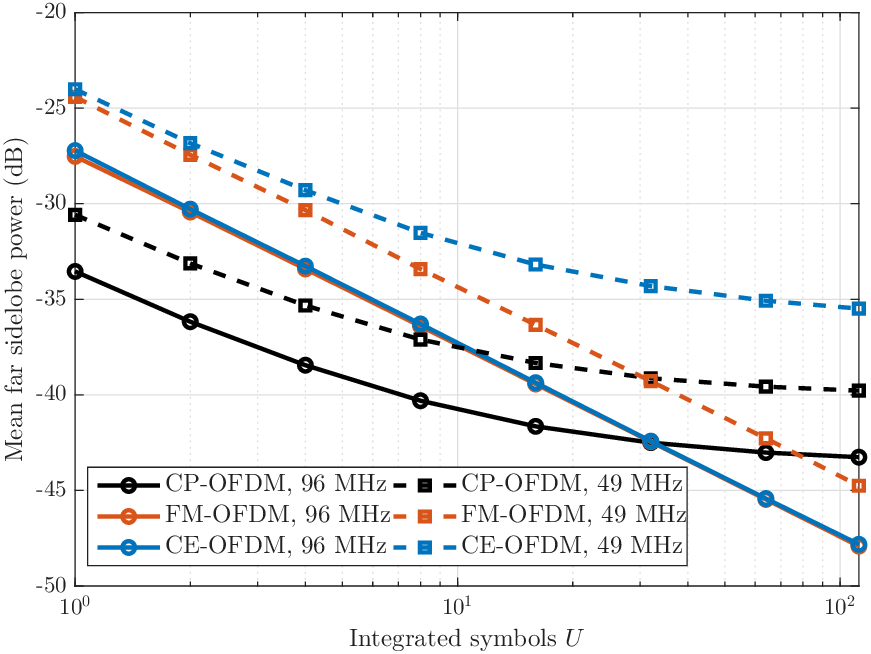}
    \caption{Mean far sidelobe power against integration length for the three waveforms.}
    \label{fig:meansll}
\end{figure}

The inset of Fig.~\ref{fig:mocc} reads the same Gaussian as an occupancy ceiling. The Nyquist edge sits at $\xi$ standard deviations of the instantaneous frequency and the $99\%$ band edge at $\eta$, so $B_{99}/f_s = \eta/\xi$ carries no waveform parameter. The measurement follows it across the sweep, stays below unity, and falls as the wrap requirement tightens.

The inset of Fig.~\ref{fig:b99} measures the occupied bandwidth against \eqref{eq:b99_law}. The measurement is above it, by $C\zeta^2$, so the excess grows as the index falls. The measured error and $C\zeta^2$ track one another up to $\zeta \simeq 1$, the limit of the expansion in \eqref{eq:R_scaled}. At the top of the sweep the measurement is below \eqref{eq:b99_law}, where the peak deviation exceeds one cycle per sample and the aliasing bound $m < 1/(2\eta)$ is reached. The measured equivalent noise bandwidth $B_e/B_{99} = 0.688$, used in Section~\ref{ssec:sim_sll} for the incoherent sidelobe floor. All three bounds of \eqref{eq:m_max} are measured, and inverting \eqref{eq:b99} lands the band on the uncorrected root $m_0 = B_{ch}/(2\eta f_s)$ surpassing it by $C\zeta^2$.

\subsection{Ambiguity Function and Sidelobe Floor}
\label{ssec:sim_sll}

The main panel of Fig.~\ref{fig:b99} measures the sidelobe floor against $m$. The coherent floor $e^{-\bar D}$ of \eqref{eq:dphi_plateau} and the incoherent term of \eqref{eq:sll} tracks the measurement over the sweep. The floor contributes nothing at $k_0 = 1$, where $\bar D$ is large, so \eqref{eq:sll_m} applies in its first branch. The dotted curve replaces the measured $B_e$ by the Gaussian $0.688\,B_{99}$. The two agree except at the bottom of the sweep, where $N_a$ rather than the deviation sets the band and the spectrum is no longer Gaussian. The operating point lies well inside the range where they agree.
\begin{figure}[!t]
    \centering
    \includegraphics[width=1\linewidth]{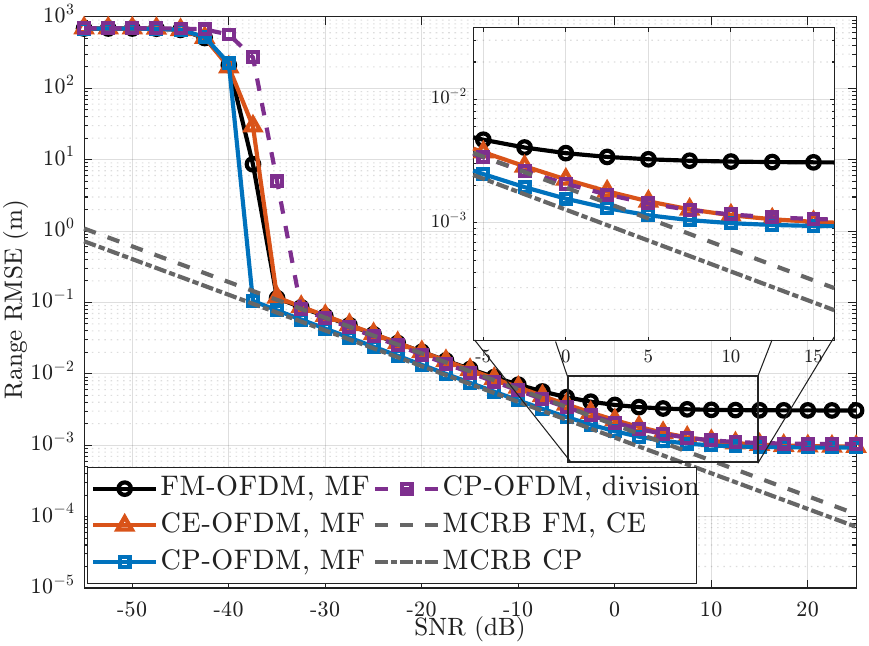}
    \caption{Range \ac{RMSE} against per-sample \ac{SNR} for the three waveforms.}
    \label{fig:sim_range}
\end{figure}

Fig.~\ref{fig:meansll} gives the far sidelobe floor against integration length at $96$ and $49$~MHz. At $U = 1$ and the wider band \ac{CP-OFDM} sits about $6$~dB below both constant-envelope waveforms, the optimality of linear modulation in average sidelobe level \cite{liu2025cp}. Integration reverses this. The \ac{FM-OFDM} floor is incoherent and falls by the full $10\log_{10}U$, \eqref{eq:sll_m} predicts at both ends of the sweep to within $0.2$~dB. The \ac{CP-OFDM} floor does not. Its mean autocorrelation away from the origin is the Dirichlet kernel of the rectangular allocation, it is identical for every data draw and survives averaging. The curves cross at $U \simeq 32$ and \ac{FM-OFDM} ends about $5$~dB lower.

The two constant-envelope waveforms agree at $96$~MHz and separate at $49$~MHz. Their coherent terms are set by different quantities. The \ac{CE-OFDM} term follows its own modulation index, so the narrower band raises it into a floor that integration cannot reach and its floor stops above \ac{FM-OFDM}. The \ac{FM-OFDM} term is $e^{-\bar D}$, this stays negligible, so that floor keeps falling and still matches \eqref{eq:sll_m}.

Raising $k_0$ removes the subcarriers that dominate \eqref{eq:dphi_plateau} and drops $\bar D$ by more than an order of magnitude. It stays far above $\ln(B_eT_u)$ across the admissible range, so $e^{-\bar D}$ stays negligible and the measured floor does not move. Within that range $k_0$ buys a \ac{DC} guard at no cost in sidelobe level.
\begin{figure}[!t]
    \centering
    \includegraphics[width=1\linewidth]{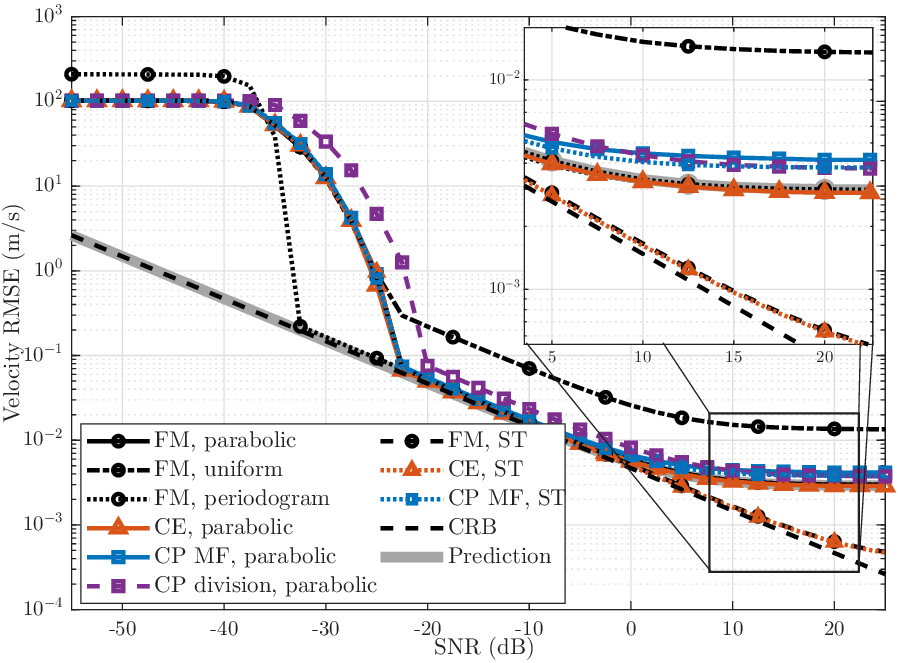}
    \caption{Velocity \ac{RMSE} against per-sample \ac{SNR}. ST is single-target}
    \label{fig:sim_velo}
\end{figure}


\subsection{Range and Velocity Estimation}
\label{ssec:sim_estimation}

Fig.~\ref{fig:sim_range} gives range \ac{RMSE} against \ac{SNR}. The three matched filters reach threshold at nearly the same \ac{SNR}, since the compressed peak is set by the frame energy and all three send the same power over the same frame. Spectrum shape does not move the threshold, we see that division receiver reaches threshold one step later due to data division.\\
Above threshold all four curves follow their bounds. The \ac{MCRB} depends on the \ac{RMS} bandwidth, not on the occupied bandwidth, the \ac{CP-OFDM} bound lies $3.6$~dB below the \ac{CE} ones. Occupancy fixes the resolution in \eqref{eq:range_resolution} and the second moment fixes the accuracy, and the division receiver stays $2.8$~dB above the \ac{CP-OFDM} matched filter, it is $10\log_{10}\mathbb{E}\big[1/|X|^2\big]$ for unit-power $16$-\ac{QAM}. Both use the same waveform, so the gap is a receiver side loss.\\
All curves flatten at high \ac{SNR}. Removing the interferer leaves the \ac{FM-OFDM} floor unchanged and lowers the \ac{CE-OFDM} one, so only the \ac{FM-OFDM} floor comes from the waveform. Therefore the floor levels do not depend on bandwidth in this case.

Fig.~\ref{fig:sim_velo} gives velocity \ac{RMSE} against \ac{SNR}. Equation \eqref{eq:optimal_variance} predicts the saturated \ac{FM-OFDM} \ac{RMSE} to $0.3$~dB with nothing fitted, The measured loss matches \eqref{eq:weight_penalty}, and both estimators reach the same floor, so the limit is in the waveform.

Removing the interferer lowers the \ac{FM-OFDM} and \ac{CE-OFDM} floors and leaves \ac{CP-OFDM} where it was. The \ac{CP-OFDM} floor is therefore set by its own data. Two terms contribute to its slow-time phase, one from the varying envelope and one from the spectral centroid of the data. A \ac{CE} removes the first and an empty \ac{DC} bin removes the second. \ac{FM-OFDM} and \ac{CE-OFDM} meet both conditions because $k_0 \ge 1$.

\subsection{Target Masking}
\label{ssec:sim_masking}
\begin{figure}[!t]
    \centering
    \includegraphics[width=0.8\linewidth]{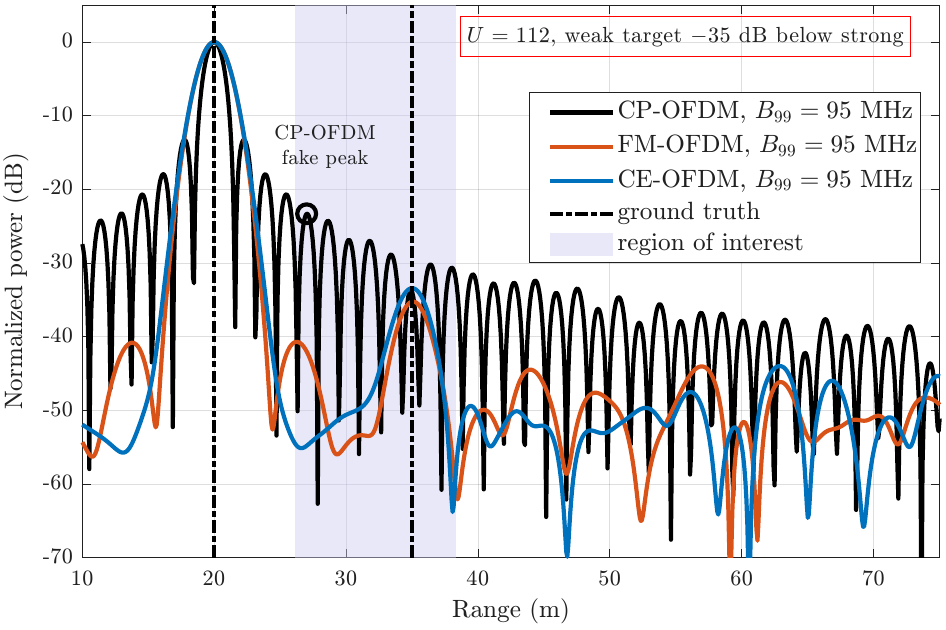}
    \caption{Zero Doppler cut of a two-target scene at equal occupied bandwidth.}
    \label{fig:zdcut}
\end{figure}

A weak scatterer is placed beside a strong one and lowered until the receiver loses it. Both targets share one velocity and the Doppler is known and removed, which isolates the range domain and gives the worst case, since any Doppler separation moves the weak target off the sidelobes of the strong one.\\
Fig.~\ref{fig:zdcut} shows the zero Doppler cut for the scenario of \cite{liu2025cp} with our waveform parameters. A weak target at $35$~m sits $35$~dB below a strong target at $20$~m, and each profile is normalized to its own peak. All three waveforms place a comparable response on the weak target, but they differ in the sidelobe level they leave around it. For \ac{FM-OFDM} and \ac{CE-OFDM} the largest response there is the weak target. For \ac{CP-OFDM} it is a sidelobe at $27$~m, standing $10.7$~dB above it.\\
The mechanism is the one separated in Section~\ref{ssec:sim_sll}. Integration drains the incoherent part of the floor and leaves the coherent part, which does not spread evenly over lags but concentrates on a few. A mean sidelobe level therefore understates what a detector sees, since the decision is set by the peak inside the gate.

This gain is not free. Matching is on $B_{99}$, and \ac{CP-OFDM} fills that band with a flat spectrum, so it has the narrowest mainlobe and the finest two target resolution of the three. The spectra of \ac{FM-OFDM} and \ac{CE-OFDM} lower the sidelobes at the cost of a wider mainlobe. The hard band edge that gives \ac{CP-OFDM} its resolution is the same edge that gives it the coherent floor.

We ran a two target detection simulation test that quantifies this masking. The weak target is swept from $0$ to $-55$~dB relative to the strong one at a per-sample \ac{SNR} of $6$~dB, so the measurement is sidelobe limited rather than noise limited. The half-detection point is $-23$~dB for \ac{CP-OFDM}, $-43$~dB for \ac{FM-OFDM} and $-44$~dB for \ac{CE-OFDM}, giving the \ac{CE} waveforms about $20$~dB more dynamic range at equal bandwidth, power and integration length.

\section{Conclusion}
\label{sec:conclusion}
This paper investigated \ac{FM-OFDM} as a \ac{CE} sensing waveform against \ac{CP-OFDM} and \ac{CE-OFDM} under matched bandwidth, with a closed-form bandwidth bound setting the modulation index. Although \ac{FM-OFDM} can attain a lower single-symbol sidelobe floor, its data-dependent sidelobe phase adds incoherently over a frame, scaling as $\sqrt{N}$, whereas \ac{CP-OFDM} sidelobes add coherently with full integration gain. After frame integration, \ac{FM-OFDM} therefore detects weaker targets. The proposed differential Doppler estimator preserves data dependent phase in the first differences of the weights, admits a closed-form variance, and estimates both the sensing floor and the crossover, which aligns with simulation results. At zero delay, the \ac{AF}  is data-independent, unlike linearly modulated waveforms. Two tradeoffs remain; the differential estimator only matches a periodogram above threshold and degrades below it, and \ac{CP-OFDM} retains a fixed range advantage governed by the \ac{RMS} bandwidth. Future work includes low-PAPR alternatives such as spectrally shaped \ac{DFT}-s-OFDM, bistatic operation, rate-sensing tradeoffs under shared bandwidth, and multi-antenna extensions.

\bibliographystyle{IEEEtran}
\bibliography{IEEEfull,References}

\end{document}